\documentstyle[twoside,fleqn,espcrc2]{article}
\newcommand{\half}{{\scriptstyle{{1\over 2}}}}
\def\beqa{\begin{eqnarray}}
\def\eeqa{\end{eqnarray}}
\def\bea{\begin{array}}
\def\eea{\end{array}}
\def\cT{{\beta}}

\def\cA{{\cal{A}}}
\def\acosh{{\rm acosh}} 
\def\diag{{\rm diag}} 
\def\tr{{\rm tr}} 
\def\Tr{{\rm Tr}} 
\def\pl{{{\cal P}_\infty}}
\def\plo{{{\cal P}_\infty^0}}

\title{\vskip-5mm 
Abelian projected monopoles - to be or not to be\thanks{
Presented at Lattice 2001, 18-24 August 2001, Berlin.
}\vskip-2.3cm\hfill\small INLO-PUB-03/01\vskip2cm
}
\author{Pierre van Baal\address{Instituut-Lorentz for Theoretical Physics, 
University of Leiden,\\PO Box 9506, NL-2300 RA Leiden, The Netherlands.}}
\begin{document}
\begin{abstract}
We analyse what happens with two merging constituent monopoles for the $SU(3)$ 
caloron. Identified through degenerate eigenvalues (the singularities or 
defects of the abelian projection) of the Polyakov loop, it follows that there 
are defects that are not directly related to the actual constituent monopoles. 
\end{abstract}
\maketitle
\section{Introduction}
Finite temperature instantons (calorons) have a rich structure if one allows
the Polyakov loop, $P(\vec x)\!=\!P\exp(\int_0^\cT\! A_0(\vec x,t) dt)$ in the 
periodic gauge $A_\mu(t,\vec x)\!=\!A_\mu(t+\cT,\vec x)$, to be non-trivial at 
spatial infinity (specifying the holonomy). It implies the spontaneous 
breakdown of gauge symmetry. For a charge one $SU(n)$ caloron, the location 
of the $n$ constituent monopoles can be identified through: 
i. Points where two eigenvalues of the Polyakov loop coincide, which is
where the $U^{n\!-\!1}(1)$ symmetry is partially restored to 
$SU(2)\!\times\!U^{n\!-\!2}(1)$. 
ii. The centers of mass of the (spherical) lumps. 
iii. The Dirac monopoles (or rather dyons, due to self-duality) as the 
sources of the abelian field lines, extrapolated back to the cores. 
If well separated and localised, all these coincide~\cite{KvB2,Dub}. Here 
we study the case of two constituents coming close together for $n\!\geq\!3$, 
with an example for $SU(3)$.

The eigenvalues of $\pl\!\equiv\!\lim_{|\vec x|\rightarrow\infty}P(\vec x)$ can
be ordered by a constant gauge transformation $W_\infty$
\beqa
&&\hskip-3mm W_\infty^\dagger\pl W_\infty\!=\!\plo\!=\!\exp[2\pi i\,\diag
(\mu_1,\ldots,\mu_n)],\nonumber\\ &&\hskip-3mm\mu_1\leq\ldots\leq\mu_n\leq
\mu_{n+1}\!\equiv\!1\!+\!\mu_1,
\eeqa
with $\sum_{m=1}^n\mu_m\!=\!0$. The constituent monopoles have masses 
$8\pi^2\nu_i$, where $\nu_i\equiv\mu_{i+1}-\mu_i$ (using the classical scale 
invariance to put the extent of the euclidean time direction to one, $\beta=
1$). In the same way we can bring $P(\vec x)$ to this form by a {\em local} 
gauge function, $P(\vec x)=W(\vec x)P_0(\vec x)W^\dagger(\vec x)$. We note 
that $W(\vec x)$ (unique up to a residual abelian gauge rotation) 
and $P_0(\vec x)$ will be smooth, except where two (or more) eigenvalues 
coincide.

The ordering shows there are $n$ different types of singularities (called 
defects~\cite{FTW}), for each of the {\em neighbouring} eigenvalues to 
coincide. The first $n-\!1$ are associated with the basic monopoles (as part
of the inequivalent $SU(2)$ subgroups related to the generators of the Cartan 
subgroup). The $n^{\rm th}$ defect arises when the first and the last 
eigenvalue (still neighbours on the circle) coincide. Its magnetic charge 
ensures charge neutrality of the caloron. The special status~\cite{KvBn,CKvB} 
of this defect also follows from the so-called Taubes winding~\cite{Tau}, 
supporting the non-zero topological charge~\cite{KvB2}.

\section{Puzzle}
To analyse the lump structure when two constituents coincide, we recall the 
simple formula for the $SU(n)$ action density~\cite{KvBn}. 
\beqa
&&\hskip-6mm\Tr F_{\mu\nu}^{\,2}(x)\!=\!\partial_\mu^2\partial_\nu^2\log\left[
\half\tr(\cA_n\cdots \cA_1)-\cos(2\pi t)\right],\nonumber\\
&&\hskip-6mm\cA_m\equiv\frac{1}{r_m}\left(\!\!\!\bea{cc}r_m\!\!&|\vec y_m\!\!
-\!\vec y_{m+1}|\\0\!\!&r_{m+1}\eea\!\!\!\right)\left(\!\!\!
\bea{cc}c_m\!\!&s_m\\s_m\!\!&c_m\eea\!\!\!\right),
\eeqa
with $\vec y_m$ the center of mass location of the $m^{\rm th}$ constituent 
monopole. We defined $r_m\!\equiv\!|\vec x\!-\!\vec y_m|$, $c_m\!\equiv\!
\cosh(2\pi\nu_m r_m)$, $s_m\!\equiv\!\sinh(2\pi\nu_m r_m)$, as well as
$\vec y_{n+1}\!\equiv\!\vec y_1$, $r_{n+1}\!\equiv\!r_1$. We are interested 
in the case where the problem of two coinciding constituents in $SU(n)$ is 
mapped to the $SU(n\!-\!1)$ caloron. For this we restrict to the case where
$\vec y_m\!=\!\vec y_{m+1}$ for some $m$, which for $SU(3)$ is {\em always} 
the case when two constituents coincide. Since now $r_m\!=\!r_{m+1}$, one 
easily verifies that $\cA_{m+1}\cA_m\!\!=\!\cA_{m+\!1}[\nu_{m+\!1}\!
\rightarrow\!\nu_m\!+\!\nu_{m+\!1}]$, describing a {\em single} constituent 
monopole (with properly combined mass), reducing eq.~(2) to the action density 
for the $SU(n\!-\!1)$ caloron, with $n\!-\!1$ constituents.

The topological charge can be reduced to surface integrals near the 
singularities with the use of $\tr(P^\dagger dP)^3\!=d~3 \tr((P_0^\dagger 
A_W P_0+2P_0^\dagger dP_0)\wedge A_W)=d~3\tr(A_W\!\wedge(2A_W\log P_0+P_0A_W 
P_0^\dagger))$, where $A_W\!\equiv W^\dagger dW$. If one assumes {\em all} 
defects are pointlike, this can be used to show that for each of the $n$ 
types the (net) number of defects has to equal the topological charge, the 
type being selected by the branch of the logarithm (associated with the $n$ 
elements in the center)~\cite{FTW}. One might expect the defects to merge 
when the constituent monopoles do. A triple degeneracy of eigenvalues for 
$SU(3)$ implies the Polyakov loop takes a value in the center. Yet this can 
be shown {\em not} to occur for the $SU(3)$ caloron with {\em unequal} masses.
We therefore seem to have (at least) one more defect than the number 
of constituents, when $\vec y_m\!\rightarrow\!\vec y_{m+1}$.

\section{Example}

We will study in detail a generic example in $SU(3)$, with $(\mu_1,\mu_2,\mu_3)
\!=\!(-17,-2,19)/60$. We denote by $\vec z_m$ the position associated with the 
$m^{\rm th}$ constituent where two eigenvalues of the Polyakov loop coincide. 
In the gauge where $\pl=\plo$ (see eq.~(1)), we established 
numerically~\cite{Dub} that 
\beqa
P_1\!\!=\!\!P(\vec z_1)\!\!=\!\!\diag(\hphantom{-}e^{-\pi i\mu_3},\hphantom{-}
                    e^{-\pi i\mu_3},\hphantom{-}e^{2\pi i\mu_3}),\nonumber\\
P_2\!\!=\!\!P(\vec z_2)\!\!=\!\!\diag(\hphantom{-}e^{2\pi i\mu_1},\hphantom{-}
                    e^{-\pi i\mu_1},\hphantom{-}e^{-\pi i\mu_1}),\\
P_3\!\!=\!\!P(\vec z_3)\!\!=\!\!\diag(-e^{-\pi i\mu_2},\hphantom{-}
                    e^{2\pi i\mu_2},-e^{-\pi i\mu_2}).\nonumber
\eeqa
This is for {\em any} choice of holonomy and constituent locations (with the 
proviso they are well separated, i.e. their cores do not overlap, in which 
case to a good approximation $\vec z_m\!=\!\vec y_m$). Here we take $\vec y_1
\!=\!(0,0,10+d)$, $\vec y_2\!=\!(0,0,10\!-\!d)$ and $\vec y_3\!=\!(0,0,-10)$. 
The limit of coinciding constituents is achieved by $d\!\rightarrow\!0$. With 
this geometry it is simplest to follow for changing $d$ the location where two
eigenvalues coincide. In very good approximation, as long as the first two 
constituents remain well separated from the third constituent (carrying 
the Taubes winding), $P_3$ will be constant in $d$ and the $SU(3)$ gauge 
field~\cite{Dub} of the first two constituents will be constant in time
(in the periodic gauge). Thus $P(\vec z_m)=\exp(A_0(\vec z_m))$ for $m=1,2$,
greatly simplifying the calculations.

When the cores of the two approaching constituents start to overlap, $P_1$ 
and $P_2$ are no longer diagonal (but still block diagonal, mixing the lower 
$2\!\times\!2$ components). At $d=0$ they are diagonal again, but $P_2$ 
will be no longer in the fundamental Weyl chamber. A Weyl reflection maps
it back, while for $d\neq0$ a more general gauge rotation back to the Cartan 
subgroup is required to do so, see fig.~1. At $d=0$, {\em each} $P_m$ (and 
$\pl$) lies on the dashed line, which is a direct consequence of the reduction 
to an $SU(2)$ caloron. 

To illustrate this more clearly, we give the expressions for $P_m$ (which we 
believe to hold for any non-degenerate choice of the $\mu_i$) when 
$d\rightarrow0$:
\beqa
\hat P_1\!\!=\!\!P(\vec z_1)\!\!=\!\!\diag(\hphantom{-}e^{2\pi i\mu_2},
        \hphantom{-}e^{2\pi i\mu_2},\hphantom{-}e^{-4\pi i\mu_2}),\!\nonumber\\
\hat P_2\!\!=\!\!P(\vec z_2)\!\!=\!\!\diag(\hphantom{-}e^{-\pi i\mu_2},
         \hphantom{-}e^{2\pi i\mu_2},\hphantom{-}e^{-\pi i\mu_2}),\!\\
\hat P_3\!\!=\!\!P(\vec z_3)\!\!=\!\!\diag(-e^{-\pi i\mu_2},
          \hphantom{-}e^{2\pi i\mu_2},-e^{-\pi i\mu_2}).\!\nonumber
\eeqa
These can be factorised as $\hat P_m=\hat P_2 Q_m$, where $\hat P_2$ describes 
an overall $U(1)$ factor. In terms of $Q_1\!=\!\diag(e^{3\pi i\mu_2},1, 
e^{-3\pi i\mu_2})$, $Q_2\!=\!\diag(1,\!1,\!1)\!=\!{\bf 1}$ and $Q_3\!=
\!\diag(-1,1,-1)$ the $SU(2)$ embedding in $SU(3)$ becomes obvious. It 
leads for $Q_2$ to the trivial and for $Q_3$ to the non-trivial element 
of the center of $SU(2)$ (appropriate for the latter, carrying the Taubes 
winding). On the other hand, $Q_1$ corresponds to $\diag(e^{3\pi i\mu_2},
e^{-3\pi i\mu_2})$, which for the $SU(2)$ caloron is not related to coinciding 
eigenvalues. For $d\!\rightarrow\!0$, fig.~2 shows that $\vec z_1$ gets 
``stuck'' at a {\em finite} distance (0.131419) from $\vec z_2$.
\begin{figure}[htb]
\vspace{3.1cm}
\includegraphics{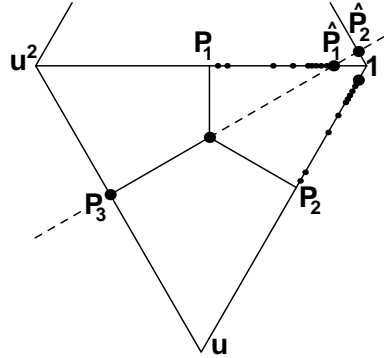}
\caption{The fundamental Weyl chamber with the positions of $P_m$ indicated 
at $d=2,$ 1, .2, .1, .05, .04, .03, .02, .01, .005, .001, .0005, and (large 
dots) 0. The perpendiculars point to $\pl$ (center), and emanate from the 
values of $P_m$ for well separated constituents. The dashed line shows the 
$SU(2)$ embedding for $d=0$. ($u\equiv\exp(2\pi i/3)\,{\bf 1}$)}
\end{figure}

\section{Resolution}

The $SU(2)$ embedding determines the caloron solution for $d\!=\!0$, with 
constituent locations $\vec y_1^{\,\prime}\!=\!\vec y_2$ and $\vec y_2^{\,
\prime}\!=\!\vec y_3$, and masses $\nu_1^{\,\prime}\!=\!\nu_1\!+\nu_2\!=\!
\mu_3\!-\!\mu_1$ and $\nu_2^{\,\prime}\!=\!\nu_3$. The best proof for the 
spurious nature of the defect is to calculate its location purely in terms 
of this $SU(2)$ caloron, by demanding the $SU(2)$ Polyakov loop to equal 
$\diag(e^{3\pi i\mu_2},e^{-3\pi i\mu_2})$. For this we can use the analytic 
expression~\cite{GGMvB} of the $SU(2)$ Polyakov loop along the $z$-axis. 
The location of the spurious defect, $\vec z_1\!=\!(0,0,z)$, is found by 
solving $3\pi\mu_2\!=\!\pi\nu_2^{\,\prime}\!-\!\half\partial_z\acosh[
\half\tr(\cA_2^{\prime}\cA_1^{\prime})]$. For our example, $z\!=\!10.131419$ 
indeed verifies this equation. 

With the $SU(2)$ embedded result at hand, we find that only for $\mu_2\!=\!0$ 
the defects merge to form a triple degeneracy. Using $3\mu_2\!=\!\nu_1\!-\!
\nu_2$, this is so for coinciding constituent monopoles of {\em equal} mass. 
For {\em unequal} masses the defect is always spurious, but it tends to stay 
within reach of the non-abelian core of the coinciding constituent monopoles, 
except when the mass difference approaches its extremal values $\pm(1\!-\!
\nu_3)$, see fig.~2 (bottom). At these extremal values one of the $SU(3)$ 
constituents becomes massless and {\em delocalised}, which we excluded for 
$d\neq0$. 
\begin{figure}[htb]
\vspace{5.7cm}
\includegraphics{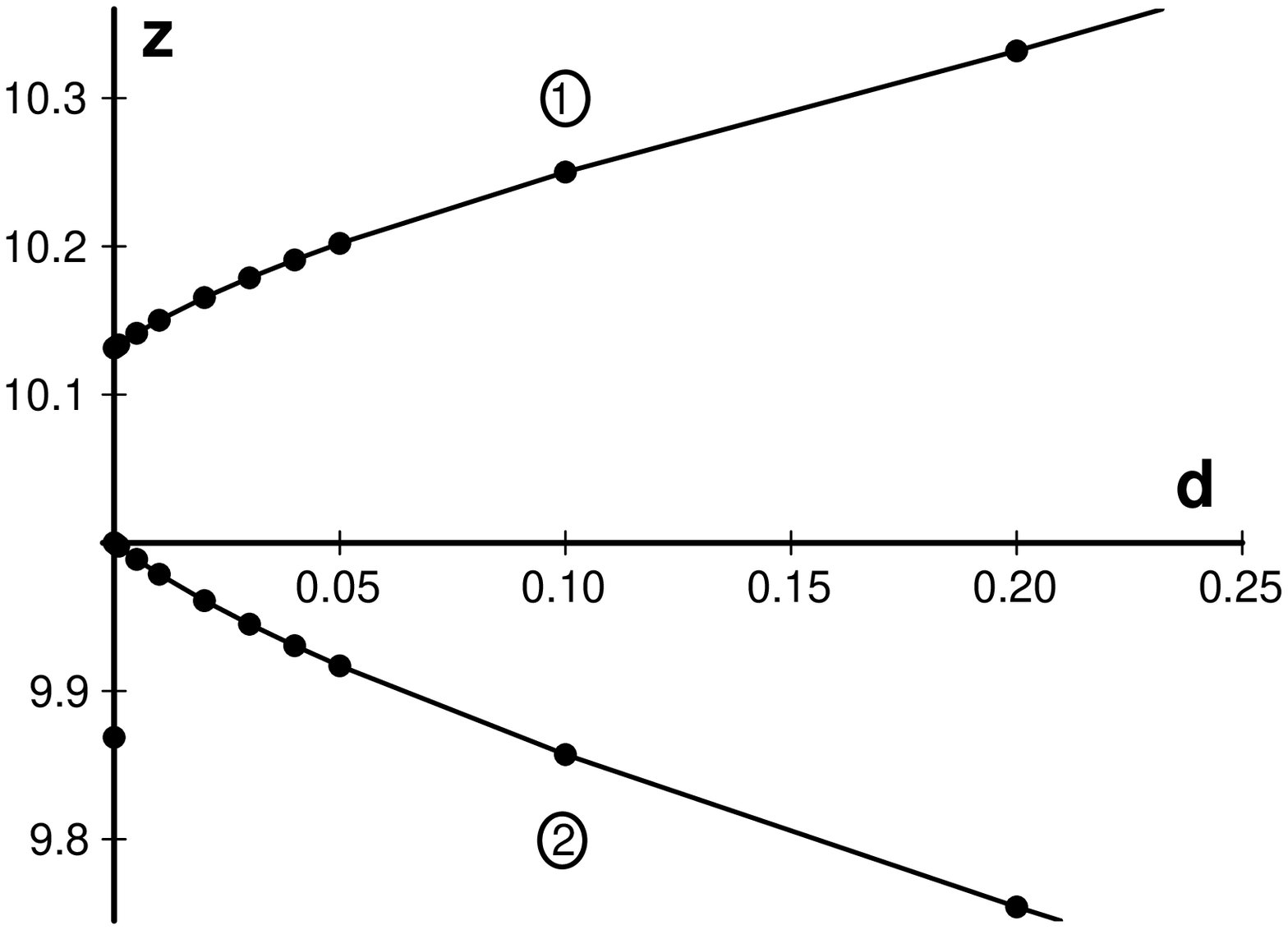}
\includegraphics{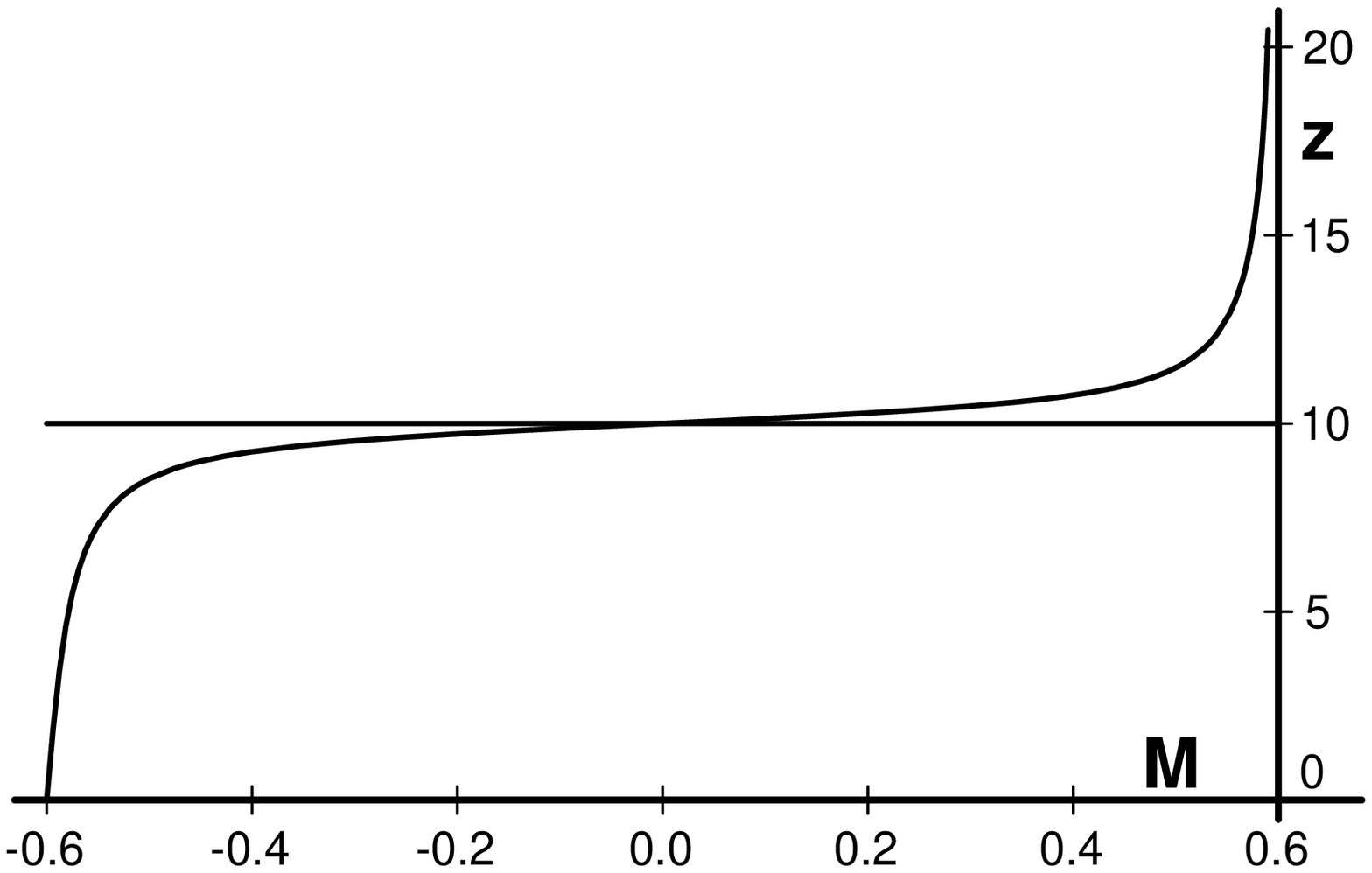}
\caption{The defect locations $\vec z_1$ and $\vec z_2$, along the $z$-axis,
for $M\!\equiv\!\nu_2\!-\!\nu_1\!=\!0.1$ as a function of $d$ (top), and for 
$d\!\rightarrow\!0$ as a function of $M$ (bottom).}
\end{figure}

However, the limit $d\!\rightarrow\!0$ is singular due to the {\em global} 
decomposition into $SU(2)\!\times\!U(1)$ at $d=0$. Gauge rotations $U$ in 
the global $SU(2)$ subgroup do not affect $\hat P_2$, and therefore any 
$UQ_1U^\dagger$ gives rise to the {\em same} accidental degeneracy. In 
particular solving $-3\pi\mu_2\!=\!\pi\nu_2^{\,\prime}\!-\!\half\partial_z
\acosh[\half\tr(\cA_2^{\prime} \cA_1^{\prime})]$ (corresponding to the Weyl 
reflection $Q_1\!\rightarrow\!Q_1^\dagger$) yields $z\!=\!9.868757$ for $\mu_2
\!=-\!1/30$ (isolated point in fig.~2 (top)). Indeed, $U\!\in\!SU(2)/U(1)$ 
traces out a (nearly spherical) {\em shell} where two eigenvalues of $P$ 
coincide (note that for $\mu_2\!=\!0$ this shell collapse to a single point, 
$z\!=\!10$). A perturbation tends to remove this accidental degeneracy. 

\section{Lesson}

Abelian projected monopoles are not always what they seem to be, even 
though required by topology. {\em Topology} cannot be localised, no matter 
how tempting this may seem for smooth fields. 

\section*{Acknowledgements}
I am grateful to Andreas Wipf for his provocative question that led to 
this work. I thank Jan Smit, and especially Chris Ford, for discussions.

\end{document}